\documentclass{article}

\PassOptionsToPackage{numbers, compress}{natbib}
\usepackage[preprint]{neurips_2025}

\usepackage[utf8]{inputenc} 
\usepackage[T1]{fontenc}    
\usepackage{hyperref}       
\usepackage{url}            
\usepackage{booktabs}       
\usepackage{amsfonts}       
\usepackage{nicefrac}       
\usepackage{microtype}      
\usepackage{xcolor}         
\usepackage{amsmath}
\usepackage{subcaption}
\usepackage{graphicx}

\title{Ordinal Label-Distribution Learning with Constrained Asymmetric Priors for Imbalanced Retinal Grading}

\author{%
  Nagur Shareef Shaik \\
  Georgia State University\\
  Atlanta, GA 30303 \\
  \texttt{nshaik3@student.gsu.edu} \\
  \And
  Teja Krishna Cherukuri \\
  Georgia State University\\
  Atlanta, GA 30303 \\
  \texttt{tcherukuri1@student.gsu.edu} \\
  \AND
  Adnan Masood \\
  UST Global Inc\\
  Aliso Viejo, CA 92656 \\
  \texttt{adnan.massod@ust.com} \\
  \And
  Ehsan Adeli \\
  Stanford University\\
  Palo Alto, CA 94305 \\
  \texttt{eadeli@stanford.edu} \\
  \And
  Dong Hye Ye \\
  Georgia State University\\
  Atlanta, GA 30303 \\
  \texttt{dongye@gsu.edu} \\
}

\begin{document}

\maketitle

\begin{abstract}
Diabetic retinopathy grading is inherently ordinal and long-tailed, with minority stages being scarce, heterogeneous, and clinically critical to detect accurately. Conventional methods often rely on isotropic Gaussian priors and symmetric loss functions, misaligning latent representations with the task’s asymmetric nature. We propose the Constrained Asymmetric Prior Wasserstein Autoencoder (CAP-WAE), a novel framework that addresses these challenges through three key innovations. Our approach employs a Wasserstein Autoencoder (WAE) that aligns its aggregate posterior with a \emph{asymmetric prior}, preserving the heavy-tailed and skewed structure of minority classes. The latent space is further structured by a \emph{Margin-Aware Orthogonality and Compactness (MAOC)} loss to ensure grade-ordered separability. At the supervision level, we introduce a direction-aware ordinal loss, where a lightweight head predicts asymmetric dispersions to generate soft labels that reflect clinical priorities by penalizing under-grading more severely. Stabilized by an adaptive multi-task weighting scheme, our end-to-end model requires minimal tuning. Across public DR benchmarks, CAP–WAE consistently achieves state-of-the-art Quadratic Weighted Kappa, accuracy, and macro-F1, surpassing both ordinal classification and latent generative baselines. t-SNE visualizations further reveal that our method reshapes the latent manifold into compact, grade-ordered clusters with reduced overlap.
\end{abstract}

\section{Introduction}
\label{sec:intro}

The steady rise of diabetes worldwide has precipitated a silent epidemic of preventable blindness, driven by \emph{diabetic retinopathy (DR)}, a microvascular complication and leading cause of vision loss among working-age adults globally \cite{li2021applications, chen2022recent}. Prolonged hyperglycemia damages retinal vasculature, progressing through stages defined by the International Clinical DR Disease Severity Scale, an ordinal taxonomy spanning five levels from no apparent retinopathy to proliferative DR \citep{wilkinson2003proposed}. Timely detection and accurate grading are critical, with estimates suggesting that up to 90\% of severe vision impairment can be prevented through routine screening and treatment \citep{chen2022recent, Ma2025ABiD}. However, manual grading is labor-intensive, prone to inter-grader variability, and difficult to scale as diabetes prevalence grows, motivating research into computer-aided DR grading from retinal fundus images \citep{he2020cabnet, castiglioni2021ai}.

Clinical deployment of automated systems faces two intertwined challenges. First, DR datasets are long-tailed: early and severe grades are underrepresented, biasing models toward majority grades and reducing sensitivity to critical minority cases \citep{bodapati2021deep, bodapati2021composite}. Second, subtle lesion-based differences between adjacent grades exacerbate misclassification, particularly under cross-center and cross-device domain shifts that degrade generalization \cite{shaik2022hinge, shaik2021lesion}. These factors highlight the need for methods that respect the \emph{ordinal} nature of DR severity while handling imbalanced label distributions and domain variability \cite{Wen2023OLDL}. Generative formulations that model latent structures aligned with disease progression offer a promising alternative to purely discriminative approaches, which often assume simplistic, symmetric class geometries \cite{Vargas2025AGAsymGrading}.

\begin{figure*}[t]
  \centering
  \begin{subfigure}[t]{0.32\linewidth}
    \centering
    \includegraphics[width=\linewidth]{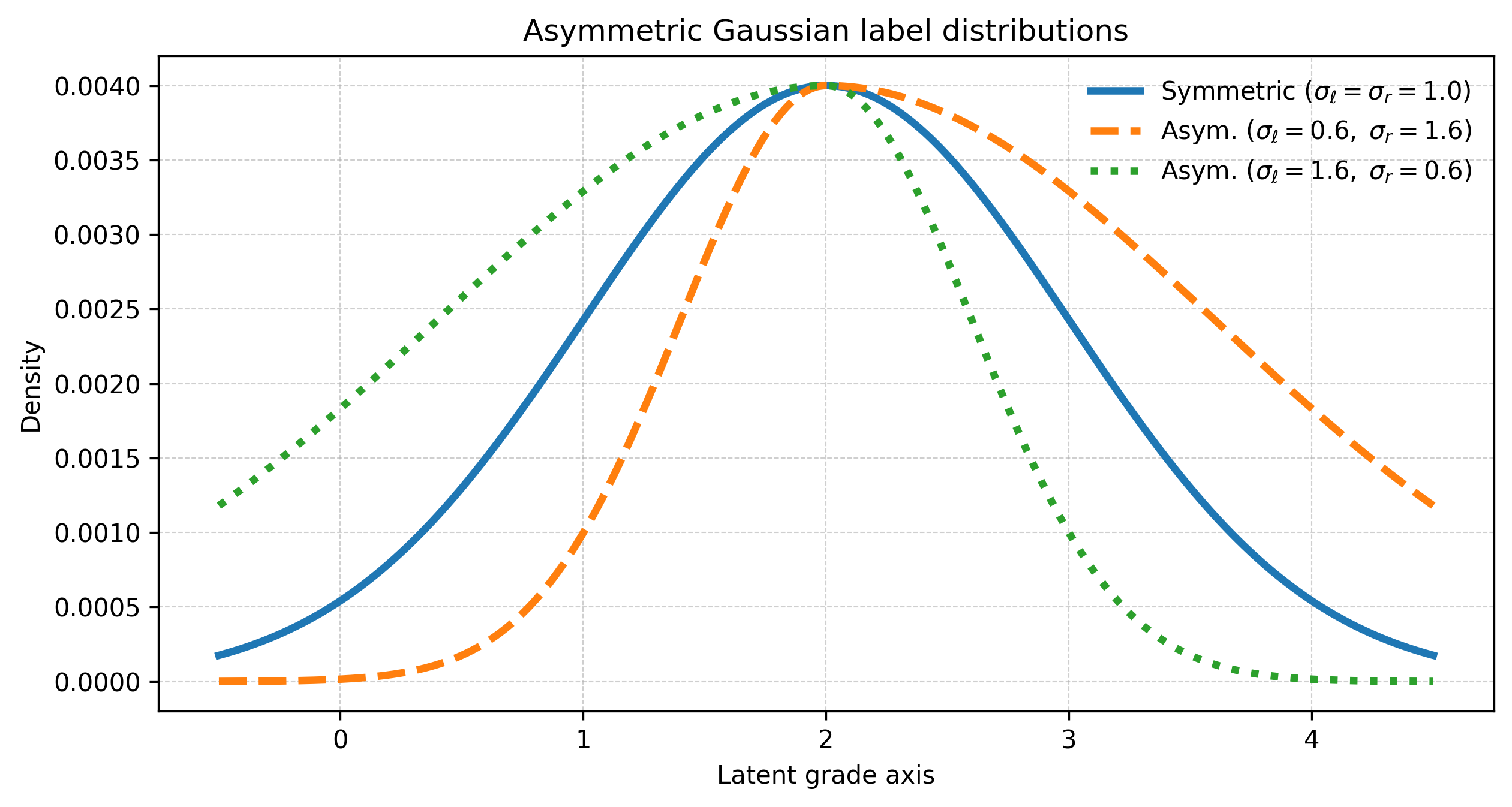}
    \caption{Asymmetric label distributions}
    \label{fig:intro-triptych:a}
  \end{subfigure}
  \hfill
  \begin{subfigure}[t]{0.67\linewidth}
    \centering
    \includegraphics[width=\linewidth]{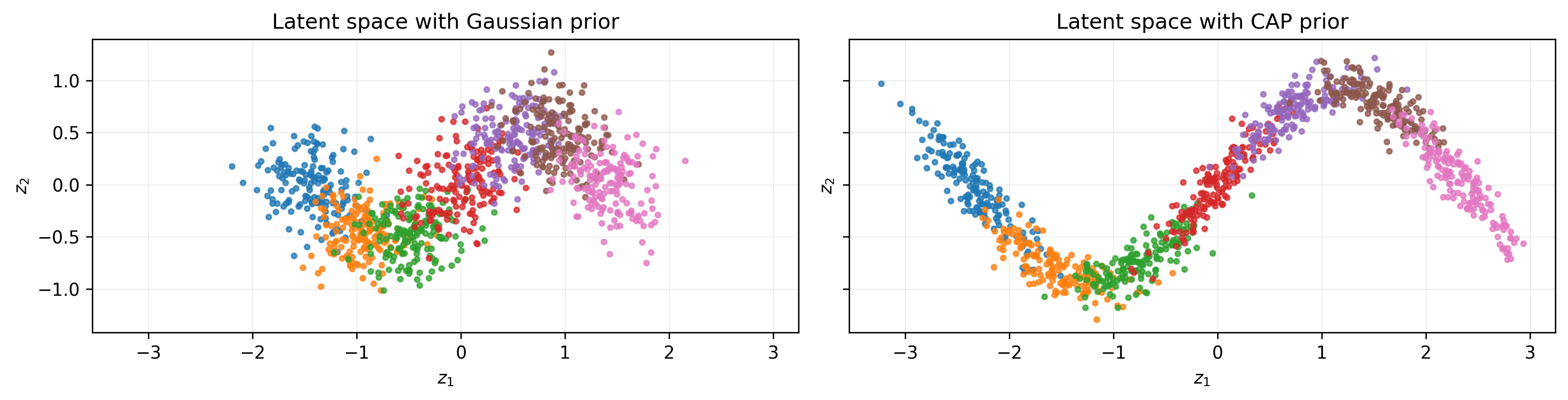}
    \caption{Latent geometry: Gaussian vs.\ CAP prior}
    \label{fig:intro-triptych:b}
  \end{subfigure}
  \vspace{-0.5ex}
  \caption{
    \textbf{(a)} Direction-aware ordinal label distributions allocate asymmetric probability mass around the reference grade.  
    \textbf{(b)} Latent manifolds: a spherical Gaussian prior (left) contracts minority modes and overlaps grades, whereas a \emph{constrained asymmetric prior} (right) preserves skew/tails and yields grade-ordered separability.}
  \label{fig:intro-triptych}
\end{figure*}

Conventional approaches mitigate imbalance with reweighting or logit adjustments and handle domain shifts with adaptation techniques \citep{lin2025g2c, Ma2025ABiD}, but they remain limited: decision boundaries are nudged toward majority grades, and one-hot supervision treats all errors equally, ignoring ordinal severity. Recent advances in \emph{ordinal label-distribution learning (OLDL)} address this by predicting distributions over ordered grades and penalizing errors by distance along the severity scale \citep{Wen2023OLDL}. Complementary studies show that asymmetric label distributions better capture clinical ambiguity than symmetric Gaussians, improving sensitivity at decision boundaries \citep{Vargas2025AGAsymGrading}, while explicit ordering priors, such as language-driven alignment, enhance robustness under distribution shifts \citep{Wang2023L2RCLIP}. These insights, illustrated in Figure~\ref{fig:intro-triptych}, suggest two gaps: (i) latent priors remain predominantly symmetric Gaussians that collapse minority structures, and (ii) supervision rarely accounts for directional clinical risks of over- vs.\ under-grading.

We propose the \emph{Constrained Asymmetric Prior Wasserstein Autoencoder} (\textbf{CAP-WAE}), the first framework to jointly integrate \emph{constrained asymmetric priors}, \emph{direction-aware ordinal supervision}, and \emph{Margin-Aware Orthogonality and Compactness} for robust DR grading. Unlike prior models that rely on symmetric Gaussian priors and one-hot supervision, CAP-WAE aligns the aggregate posterior to a skew- and tail-aware prior, preserving minority structures in the latent space. At the supervision level, it employs ordinal label distributions that penalize under-grading more severely, while adaptive multi-task weighting ensures stable, tuning/light training. Together, these innovations yield grade-ordered, well-separated manifolds and clinically aligned predictions that generalize across imbalanced and heterogeneous datasets.

The CAP-WAE framework is designed to address the following:
\begin{itemize}
  \item \textbf{Constrained asymmetric prior with Margin-Aware Orthogonality and Compactness.}  
  We introduce a Wasserstein autoencoder that aligns the aggregate posterior to a skew- and tail-aware \emph{constrained asymmetric prior}, while enforcing geometric regularization to preserve minority structures and yield grade-ordered, well-separated manifolds.
  \item \textbf{Direction-aware ordinal supervision with tuning-light training.}  
  We couple direction-aware ordinal label distributions with adaptive multi-task weighting, enabling clinically aligned grading and robust generalization without extensive hyperparameter tuning.
\end{itemize}

\section{Preliminaries}
\label{sec:preliminaries}

\paragraph{Notation.}
Let $\mathcal{Y}=\{0,1,\dots,C-1\}$ denote ordered DR grades and $\Delta^{C-1}$ the probability simplex. We use $\mathbb{I}[\cdot]$ for indicator functions and write $q_\theta(x)\in\Delta^{C-1}$ for predicted label distributions.

\subsection{Ordinal Label-Distribution Learning (OLDL)}
Label Distribution Learning models a target distribution $d\in\Delta^{C-1}$ rather than a one-hot label and minimizes a divergence $\mathcal{D}(d,q_\theta(x))$; OLDL additionally encodes \emph{order-awareness} so that larger grade gaps are penalized more strongly \citep{Wen2023OLDL}.

\paragraph{Order-aware divergences.}
Two widely used families are:
(i) \emph{Cumulative} distances on the discrete line \cite{sakai2022variants}. With $F_d(k)=\sum_{i\le k} d_i$ and $F_q(k)=\sum_{i\le k} q_i$,
\begin{equation}
\label{eq:cad}
\mathrm{CAD}(d,q)=\sum_{k=0}^{C-1}\big|F_d(k)-F_q(k)\big|,
\end{equation}
which coincides with the $L_1$ distance between discrete CDFs and is equivalent to the 1-Wasserstein distance on $\mathcal{Y}$.
(ii) \emph{Quadratic-form} distances that weight errors by grade gaps:
\begin{equation}
\label{eq:qfd}
\mathrm{QFD}(d,q)=(d-q)^\top L (d-q),\qquad
L=\mathrm{Diag}(W\mathbf{1})-W,\ \ W_{ij}=\omega(|i-j|),
\end{equation}
where $L$ is the graph Laplacian induced by a nonnegative kernel $\omega$ (e.g., linear or quadratic in $|i-j|$) \citep{Wen2023OLDL}.
OLDL variants also use Jensen–Shannon-type divergences on cumulative distributions, improving rank-sensitive metrics.

\subsection{Asymmetric Label Distributions in Medical Grading}
Clinical ambiguity is often \emph{directional}: the risk of over- vs.\ under-grading differs near boundaries. Medical imaging models therefore replace symmetric Gaussians with \emph{asymmetric}, instance-wise distributions by predicting left/right spreads \citep{Vargas2025AGAsymGrading}. A standard two-sided Gaussian parameterization is
\begin{equation}
\label{eq:asym_label}
p(j\,|\,\mu,\sigma_\ell,\sigma_r)\ \propto\
\exp\!\Bigg(
-\frac{(j-\mu)^2}{
2\big(\sigma_\ell^2\,\mathbb{I}[j<\mu]+\sigma_r^2\,\mathbb{I}[j\ge\mu]\big)+\varepsilon}
\Bigg),\quad j\in\mathcal{Y},
\end{equation}
with $(\sigma_\ell,\sigma_r)$ predicted per instance. This \emph{direction-aware} spread calibrates decision boundaries and improves sensitivity under class imbalance and label ambiguity \citep{Vargas2025AGAsymGrading, Ma2025ABiD}.

\subsection{Latent Distribution Learning}
\paragraph{VAEs.}
Variational Autoencoders maximize the ELBO
\begin{equation}
\label{eq:elbo}
\mathcal{L}_{\mathrm{ELBO}}
=\mathbb{E}_{q_\phi(z|x)}[\log p_\theta(x|z)]
-\mathrm{KL}\!\left(q_\phi(z|x)\,\|\,p(z)\right),
\end{equation}
typically with $p(z)=\mathcal{N}(0,I)$ \citep{Kingma2014VAE,Rezende2014Stochastic}. The \emph{pointwise} KL term pushes every posterior $q_\phi(z|x)$ toward the spherical prior; under long-tailed regimes this can suppress non-Gaussian structure characteristic of minority classes.

\paragraph{WAEs and aggregate alignment.}
Wasserstein Autoencoders align the \emph{aggregate} posterior $q_\phi(z)=\mathbb{E}_{p(x)}[q_\phi(z|x)]$ to a chosen prior $p(z)$ via an optimal-transport-style penalty \citep{Tolstikhin2018WAE}:
\begin{equation}
\label{eq:wae}
\mathcal{L}_{\mathrm{WAE}}
=\mathbb{E}_{p(x)}\mathbb{E}_{q_\phi(z|x)}\!\big[\ell(x,\tilde x)\big]
+\lambda\,\mathcal{D}\!\big(q_\phi(z),p(z)\big),
\quad \tilde x\sim p_\theta(x|z),
\end{equation}
instantiated with Maximum Mean Discrepancy (MMD) using a characteristic kernel $k$ \citep{Gretton2012MMD}. Given samples $\{z_i\}_{i=1}^{n}\!\sim q$ and $\{\tilde z_j\}_{j=1}^{m}\!\sim p$, a standard (biased) estimator is
\begin{equation}
\label{eq:mmd}
\mathrm{MMD}^2(q,p)
=\frac{1}{n(n-1)}\!\!\sum_{i\neq j}k(z_i,z_j)
+\frac{1}{m(m-1)}\!\!\sum_{i\neq j}k(\tilde z_i,\tilde z_j)
-\frac{2}{nm}\sum_{i,j}k(z_i,\tilde z_j).
\end{equation}
MMD-based aggregate matching is stable, sample-driven, and accommodates non-Gaussian/heavy-tailed priors, crucial for preserving minority structure in imbalanced settings. We adopt the same machinery in §\ref{sec:method} (see Eq.~\eqref{eq:mmd}) for aligning to an asymmetric prior.

\subsection{Asymmetric Generalized Gaussian (AGGD): Conceptual Background}
To model skew and tail-heaviness in latents, we leverage the \emph{asymmetric generalized Gaussian} family, which augments the generalized Gaussian with a location $\mu$, a shape (tail) parameter $\beta$, and distinct left/right scales $(\alpha_\ell,\alpha_r)$ \cite{elguebaly2013finite}. Intuitively, $\beta$ tunes tail thickness (smaller $\beta$ $\Rightarrow$ heavier tails), while $(\alpha_\ell,\alpha_r)$ control asymmetry around $\mu$. We provide the formal pdf, moments and use an AGGD sampler to draw prior samples for MMD alignment.

\section{Methodology}
\label{sec:method}

Our goal is to develop a robust framework for automated diabetic retinopathy (DR) grading that addresses long-tailed class distributions, ordinal severity scales, and clinical domain shifts. We propose the \emph{Constrained Asymmetric Prior Wasserstein Autoencoder} (CAP-WAE), which integrates generative latent modeling with order-sensitive supervision and explicit geometric regularization. We formulate retinal grading as an ordinal multi-class problem on the ordered labels $\mathcal{Y}=\{0,1,\dots,C{-}1\}$. Given an image $x\!\in\!\mathbb{R}^{H\times W\times 3}$ with grade $y\!\in\!\mathcal{Y}$, our model learns: (i) a latent representation $z$ whose aggregate distribution matches a data-informed asymmetric prior, preserving minority class structure; and (ii) an order-sensitive predictor trained with direction-aware supervision to produce clinically aligned and robust predictions.

\subsection{Architecture}

The CAP-WAE architecture consists of a deterministic encoder-decoder backbone with three lightweight heads operating on the shared latent representation.

\paragraph{Encoder-Decoder.} The encoder $f_{\phi}:\mathbb{R}^{H\times W\times 3}\to\mathbb{R}^{d}$ is a VGG16 backbone (pre-trained on ImageNet) \cite{simonyan2014very} followed by two fully connected layers, producing a latent vector $z=f_{\phi}(x)$. The decoder $g_{\theta}:\mathbb{R}^{d}\to\mathbb{R}^{H\times W\times 3}$ uses a symmetric structure of transposed convolutions to reconstruct the input image $\tilde{x}=g_{\theta}(z)$.

\paragraph{Prediction Heads.} Three shallow MLP heads are attached to the latent vector $z$: a classifier head $h_{\psi_{\mathrm{cls}}}:\mathbb{R}^{d}\to\mathbb{R}^{C}$ that outputs logits $\ell$ for class prediction; an Asymmetric-Gaussian (AG) head $a_{\psi_{\mathrm{ag}}}:\mathbb{R}^{d}\to\mathbb{R}^{2}$ that predicts left/right log-dispersions $(\log\sigma_{\ell}, \log\sigma_{r})$ used to construct direction-aware soft labels \cite{Vargas2025AGAsymGrading}; and an ordinal regression (ORM) head $r_{\psi_{\mathrm{orm}}}:\mathbb{R}^{d}\to\mathbb{R}$ that predicts a scalar severity score $s$ aligned with the grade index \cite{tang2023disease}.

\subsection{Constrained Asymmetric Prior}
\label{sec:cap}

To capture the skewed, heavy-tailed nature of features in long-tailed medical datasets, we replace the standard spherical Gaussian prior with an \emph{Asymmetric Generalized Gaussian Distribution (AGGD)}. The "constrained" aspect of our prior, which we term $p_{\text{cap}}$, refers to the fact that its parameters are fixed and empirically determined from the data distribution, providing a stable, informative target for latent space alignment.

\paragraph{Prior Definition (AGGD).}
We use a factorized prior $p_{\text{cap}}(z)=\prod_{j=1}^{d} p_{\text{cap}}(z_j)$, where each coordinate follows an AGGD with parameters $\eta_j = (\mu_j, \beta_j, \alpha_{\ell,j}, \alpha_{r,j})$. The coordinate-wise pdf is:
\begin{equation}
\label{eq:aggd-pdf}
p_{\text{cap}}(u;\eta)
=\frac{\beta}{(\alpha_{\ell}+\alpha_{r})\,\Gamma(1/\beta)}\times
\begin{cases}
\exp\!\big(-\!\big(\tfrac{|\mu-u|}{\alpha_{\ell}}\big)^{\beta}\big), & u<\mu,\\[4pt]
\exp\!\big(-\!\big(\tfrac{|u-\mu|}{\alpha_{r}}\big)^{\beta}\big), & u\ge \mu.
\end{cases}
\end{equation}
Here, $\mu$ is the location (center), $\beta$ controls tail heaviness ($\beta{=}2$ Gaussian, $\beta{=}1$ Laplace, smaller $\beta$ heavier-tailed), and $\alpha_\ell,\alpha_r$ are left/right scales governing asymmetry. Given latent vectors $\{z_i\}_{i=1}^N$, for each coordinate $j$,:
\[
\mu_j = \tfrac{1}{N}\sum_i z_{i,j}, \qquad
\alpha_{\ell,j} = \sqrt{\tfrac{1}{N_\ell}\!\sum_{i:z_{i,j}<\mu_j} (\mu_j-z_{i,j})^2}, \qquad
\alpha_{r,j} = \sqrt{\tfrac{1}{N_r}\!\sum_{i:z_{i,j}\ge\mu_j} (z_{i,j}-\mu_j)^2},
\]
where $N_\ell,N_r$ are counts on each side of $\mu_j$.

\paragraph{Aggregate Alignment via MMD.}
We align the aggregate posterior $q_\phi(z)=\mathbb{E}_{p(x)}[q_\phi(z|x)]$ to our fixed prior $p_{\text{cap}}$ using the Maximum Mean Discrepancy (MMD) penalty, as defined in Eq.~\eqref{eq:mmd}. This distribution-level matching allows the model to preserve the non-Gaussian structures of minority classes, which would otherwise be lost under the instance-level KL divergence penalty in a standard VAE.

\subsection{Order-Sensitive Supervision}
To leverage the ordinal nature of DR grades, our supervision combines three complementary loss terms.

\paragraph{Hard Classification.} A standard cross-entropy loss on the one-hot target $y$:
\begin{equation}
\label{eq:ce}
\mathcal{L}_{\mathrm{CE}} = -\log \big( \mathrm{softmax}_y(\ell) \big), \quad \text{where } \ell=h_{\psi_{\mathrm{cls}}}(z).
\end{equation}

\paragraph{Asymmetric Gaussian Soft Labels (AG-soft).} The AG head predicts instance-wise dispersions $(\sigma_\ell, \sigma_r) = \exp(a_{\psi_{\mathrm{ag}}}(z))$, which are clamped to a stable range $[\sigma_{\min}, \sigma_{\max}]$. These define a direction-aware target distribution:
\begin{equation}
\label{eq:agsoft}
p_{\mathrm{AG}}(j \mid y,\sigma_\ell,\sigma_r)\;\propto\;
\exp\!\left(
-\frac{(j-y)^2}{2\big(\sigma_\ell^2\,\mathbb{I}[j<y]+\sigma_r^2\,\mathbb{I}[j>y]+\sigma_{\mathrm{mid}}^2\,\mathbb{I}[j=y]\big)+\varepsilon}
\right),
\end{equation}
where $\sigma_{\mathrm{mid}} = (\sigma_\ell + \sigma_r)/2$ ensures a well-defined peak. This target is used in a soft cross-entropy loss, penalizing under- or over-grading differently based on the learned dispersions:
\begin{equation}
\label{eq:softce}
\mathcal{L}_{\mathrm{AG}} 
= -\sum_{j=0}^{C-1} p_{\mathrm{AG}}(j \mid y,\sigma_\ell,\sigma_r) \log \big( \mathrm{softmax}_j(\ell) \big).
\end{equation}

\paragraph{Ordinal Regression (ORM).} The ORM head predicts a scalar score $s=r_{\psi_{\mathrm{orm}}}(z)$, which is trained to align with the integer grade $y$ using a Huber loss:
\begin{equation}
\label{eq:orm}
\mathcal{L}_{\mathrm{ORM}}
= \mathrm{Huber}_{\tau}\!\big(s-y\big).
\end{equation}

\subsection{Latent Geometry Regularization}
To enforce a structured latent space with well-separated and compact class clusters, we introduce a single geometric loss term, the \textbf{Margin-Aware Orthogonality and Compactness (MAOC)} loss. This loss combines two principles from prototype-based learning. Let $\mu_c$ be the running mean of latent vectors for class $c$, and $\hat{\mu}_c = \mu_c / \|\mu_c\|_2$.
\begin{equation}
\label{eq:maoc}
\mathcal{L}_{\mathrm{MAOC}}
= \underbrace{\frac{1}{|\mathcal{P}|}\sum_{c\neq c'} \!\left[\max\!\big(0,\,\hat{\mu}_c^\top \hat{\mu}_{c'}-\delta\big)\right]^2}_{\text{Margin-Aware Orthogonality}}
+ \gamma_{\mathrm{cmp}} \cdot \underbrace{\frac{1}{N}\sum_{i=1}^{N}\|z_i-\mu_{y_i}\|_2^2}_{\text{Intra-Class Compactness}},
\end{equation}
where $\mathcal{P}$ is the set of class pairs, $\delta$ is an angular margin, and $\gamma_{\mathrm{cmp}}$ is a hyperparameter balancing the two components. This encourages class prototypes to be nearly orthogonal while pulling individual samples toward their respective class centers.

\subsection{Complete Objective and Training}
The full CAP-WAE model is trained end-to-end by minimizing a composite objective. To balance the multiple supervised loss terms without extensive hyperparameter tuning, we use an uncertainty-based adaptive weighting scheme. Let $\Theta$ denote all network parameters and $s \in \mathbb{R}^3$ be learnable log-variance parameters for weighting. The final objective is:
\begin{align}
\label{eq:capwae-total-revised}
\min_{\Theta, s} \quad \mathbb{E}_{(x,y)\sim p_{\mathrm{data}}}\Big[
&\underbrace{\|\ g_\theta(f_\phi(x)) - x\ \|_2^2}_{\mathcal{L}_{\mathrm{recon}}}
\;+\; \lambda_{\mathrm{reg}}\underbrace{\mathrm{MMD}^2\!\big(q_\phi(z),\,p_{\text{cap}}\big)}_{\mathcal{L}_{\mathrm{reg}}} \notag \\
&+ \lambda_{\mathrm{maoc}}\underbrace{\mathcal{L}_{\mathrm{MAOC}}(Z)}_{\text{Geometric Regularization}}
\;+\; \underbrace{\sum_{k \in \{\text{CE, AG, ORM}\}} \left( e^{-s_k}\mathcal{L}_k + s_k \right)}_{\text{Adaptive Ordinal Supervision}}
\Big],
\end{align}
where $z=f_\phi(x)$, $Z$ represents the batch of latent vectors, and the $\mathcal{L}_k$ are defined in Eqs.~\eqref{eq:ce}, \eqref{eq:softce}, and \eqref{eq:orm}. The hyperparameters $\lambda_{\mathrm{reg}}$ and $\lambda_{\mathrm{maoc}}$ balance the generative and geometric regularizers against the reconstruction and supervised losses. The model is trained using the AdamW optimizer with gradient clipping to ensure stability.

\section{Experimental Results}
\label{sec:results}

\subsection{Datasets}

We conduct a comprehensive evaluation on four public DR datasets to test CAP-WAE's ordinal grading performance and robustness to real-world domain shifts. These corpora utilize two distinct clinical grading taxonomies and differ in camera hardware (Zeiss, Kowa, Topcon) and patient populations, providing a rigorous testbed for generalization. The primary taxonomy is a fine-grained 7-grade scale used by Zenodo-DR-7 \cite{benitez2021dataset}, which includes C$_0$: No DR, C$_1$: Mild Non-Proliferative DR (NPDR), C$_2$: Moderate NPDR, C$_3$: Severe NPDR, C$_4$: Very Severe NPDR, C$_5$: Proliferative DR (PDR), and C$_6$: Advanced PDR. The other datasets adhere to the more common international standard 5-grade scale: 0: No DR, 1: Mild NPDR, 2: Moderate NPDR, 3: Severe NPDR, and 4: Proliferative DR (PDR). Key characteristics of each dataset are summarized in Table \ref{tab:datasets}.
\begin{table}[!ht]
  \centering
  \caption{Summary of diabetic retinopathy datasets used for evaluation.}
  \label{tab:datasets}
  \small
  \setlength{\tabcolsep}{8pt}
  \begin{tabular}{@{} l l c r r r @{}}
    \toprule
    \textbf{Dataset} & \textbf{Camera system} & \textbf{\# Classes} & \textbf{Train} & \textbf{Test} & \textbf{Total} \\
    \midrule
    Zenodo-DR-7 \cite{benitez2021dataset}  & Zeiss Visucam 500  & 7 & 605   & 152  & 757   \\
    IDRiD \cite{porwal2018indian}        & Kowa VX-10a        & 5 & 413   & 103  & 516   \\
    APTOS-2019 \cite{aptos2019}    & Heterogeneous      & 5 & 2{,}929 & 733  & 3{,}662 \\
    Messidor-2 \cite{abramoff2013automated}   & Topcon TRC-NW6     & 5 & 1{,}398 & 350  & 1{,}748 \\
    \bottomrule
  \end{tabular}
\end{table}
\paragraph{Zenodo-DR-7.} Our primary benchmark, featuring a fine-grained 7-level severity scale and an explicit long-tail distribution, making it ideal for evaluating rare-stage separability. We use the official train/test split.
\paragraph{IDRiD.} Sourced from an Indian population, this dataset allows us to assess generalization to a different camera system and demographic. We use the official 5-grade labels and data split.
\paragraph{APTOS-2019.} A large-scale dataset captured under varied clinical conditions, serving as a standard benchmark for model robustness against noise and acquisition variability.
\paragraph{Messidor-2.} A widely used clinical benchmark that is crucial for evaluating cross-device generalization from the Zeiss and Kowa systems used in the other datasets.

\subsection{Implementation Details}

All models share one PyTorch pipeline for fairness. Images are resized to \(224\times224\) (bicubic), normalized to ImageNet mean/std, and passed to an ImageNet-initialized VGG16 encoder; a symmetric transposed-conv decoder reconstructs \(\tilde{x}\). Three lightweight heads on the latent \(z\) output (i) logits, (ii) AG-soft dispersions \((\sigma_\ell,\sigma_r)\), and (iii) an ordinal score. Evaluation reports Accuracy, macro-F1 and Quadratic Weighted Kappa (QWK). Hyperparameters: latent dimension \(512\), batch size \(32\), epochs \(100\); AdamW with learning rate \(1\!\times\!10^{-4}\) and weight decay \(1\!\times\!10^{-5}\). All hyperparameters were selected empirically and tuned over multiple runs to balance reconstruction fidelity, latent regularization, and downstream predictive performance. The objective is the sum of pixel MSE (reconstruction), MMD (prior alignment), MAOC (geometry), and supervised terms (CE, AG-soft CE with softly clamped \(\sigma_\ell,\sigma_r\), Smooth-L1 ORM). Loss weights: \(\lambda_{\text{reg}}{=}0.1\), \(\lambda_{\text{MAOC}}{=}0.05\); if adaptive weighting is enabled, CE/AG/ORM are combined with learned log-variances, else fixed weights \(\{1.0, 1.0, 0.5\}\) are used. The asymmetric prior sampler uses \(\beta{=}1.2\) (tail-heaviness); MAOC uses margin \(0.1\) and compactness \(0.5\); AG-soft clamps \(\sigma\in[0.2,5.0]\). Optimization uses AMP, grad-norm clip \(1.0\), and \texttt{ReduceLROnPlateau} on validation QWK (factor \(0.2\), patience \(7\)); the best QWK checkpoint is reported.

\subsection{Quantitative Evaluation}

The comparative study in Table~\ref{tab:sota_main} highlights three key observations. 
First, conventional imbalance-aware classifiers (CE, Focal, LDAM-DRW) improve balanced accuracy, but struggle to respect the ordinal continuum of DR severity, leading to inconsistent macro-F1 across datasets. Ordinal-specific baselines (CORN, OLDL) mitigate label noise by leveraging rank information, achieving stronger QWK and F1, though still limited by rigid discriminative training. Second, latent generative approaches (VAE–KL, WAE–MMD) provide more stable calibration and smoother feature distributions, but their reliance on symmetric priors or loose Wasserstein alignment constrains separation in long-tailed classes. Third, attention- and discriminator-based models (e.g., GCG, AGDGN) report competitive accuracy but underperform in F1 on minority and severe classes, reflecting limited ordinal inductive bias.

CAP-WAE consistently outperforms all baselines across four benchmarks, with gains most pronounced on Zenodo-DR-7 (QWK $+3.2$ over AGDGN+OLDL) and IDRiD (macro-F1 $+2.6$ over the strongest baseline). The improvements stem from the synergy of generative alignment, asymmetric priors, and ordinal/geometry-aware supervision: Wasserstein matching yields smoother latent coverage; the asymmetric prior reduces bias toward majority grades; AG-Soft and ORM explicitly enforce ordinal calibration; and the proposed MAOC prior reshapes the latent space into severity-ordered, nearly orthogonal bands, improving both separability and robustness to class imbalance.

\begin{table*}[!t]
\small
\centering
\caption{\textbf{Main comparisons} on four DR benchmarks. Quadratic Weighted Kappa (QWK), Accuracy (Acc, \%), and Macro-F1 (F1, \%). †: values reported by original papers; Best in \textbf{bold}; Methods are grouped as imbalance/ordinal baselines, latent generative baselines, attention/gating models, and discriminator-based grading networks for clarity}
\label{tab:sota_main}
\setlength{\tabcolsep}{3pt}
\begin{tabular}{lccc|ccc|ccc|ccc}
\toprule
& \multicolumn{3}{c}{\textbf{Zenodo-DR-7} (7)} & \multicolumn{3}{c}{\textbf{APTOS-2019} (5)} & \multicolumn{3}{c}{\textbf{Messidor-2} (5)} & \multicolumn{3}{c}{\textbf{IDRiD} (5)} \\
\cmidrule(lr){2-4}\cmidrule(lr){5-7}\cmidrule(lr){8-10}\cmidrule(lr){11-13}
Method & QWK & Acc & F1 & QWK & Acc & F1 & QWK & Acc & F1 & QWK & Acc & F1 \\
\midrule
CE (balanced)        & 0.83 & 86.10 & 83.00 & 0.82 & 82.50 & 76.20 & 0.81 & 81.00 & 76.80 & 0.78 & 78.80 & 74.20 \\
Focal ($\gamma{=}2$) & 0.85 & 87.20 & 84.10 & 0.83 & 83.80 & 77.70 & 0.82 & 82.00 & 77.80 & 0.79 & 79.80 & 75.10 \\
Logit-Adj            & 0.86 & 87.90 & 84.80 & 0.84 & 84.90 & 78.90 & 0.83 & 82.80 & 78.60 & 0.80 & 80.60 & 75.90 \\
LDAM-DRW \cite{cao2019ldam}             & 0.87 & 88.50 & 85.60 & 0.85 & 84.60 & 79.80 & 0.84 & 83.50 & 79.40 & 0.81 & 81.50 & 76.80 \\
CORN \cite{cao2021corn}                & 0.88 & 89.10 & 86.20 & 0.86 & 84.20 & 80.40 & 0.85 & 84.10 & 80.10 & 0.82 & 82.30 & 77.60 \\
OLDL (S) \cite{Wen2023OLDL}          & 0.89 & 89.70 & 86.80 & 0.87 & 84.70 & 81.10 & 0.86 & 84.60 & 80.70 & 0.83 & 82.90 & 78.30 \\
OLDL (AS) \cite{Wen2023OLDL} & 0.90 & 90.00 & 87.50 & 0.88 & 85.90 & 82.30 & 0.87 & 85.30 & 81.60 & 0.84 & 83.80 & 79.10 \\
\midrule
VAE–KL \cite{Kingma2014VAE}        & 0.84 & 86.70 & 83.50 & 0.83 & 83.40 & 77.90 & 0.82 & 82.20 & 77.40 & 0.79 & 80.20 & 75.40 \\
WAE–MMD \cite{Tolstikhin2018WAE}       & 0.86 & 88.00 & 84.90 & 0.85 & 85.00 & 79.50 & 0.84 & 83.60 & 79.00 & 0.81 & 81.60 & 76.50 \\
\midrule
ViT$^{\dagger}$ \cite{dosovitskiy2021vit}      & --   & 84.61 & 83.19 & --   & 83.22 & 67.83 & --   & 76.79 & 61.47 & --   & 61.17 & 46.18 \\
GCG$^{\dagger}$ \cite{cherukuri2024guided}      & 0.931 & 90.13 & 88.49 & --   & 85.29 & 70.57 & --   & 80.23 & 73.85 & --   & 72.14 & 68.34 \\
\midrule
DGN \cite{Vargas2025AGAsymGrading}    & 0.87 & 88.30 & 85.20 & 0.84 & 84.60 & 78.40 & 0.83 & 83.00 & 78.80 & 0.80 & 80.80 & 76.20 \\
AGDGN \cite{Vargas2025AGAsymGrading} & 0.89 & 89.50 & 86.50 & 0.86 & 86.10 & 80.10 & 0.85 & 84.30 & 80.40 & 0.82 & 82.10 & 77.80 \\
AGDGN+OLDL   & 0.90 & 89.90 & 87.20 & 0.87 & 86.80 & 81.50 & 0.86 & 84.90 & 81.10 & 0.83 & 82.70 & 78.60 \\
\midrule
\textbf{CAP-WAE}   & \textbf{0.94} & \textbf{91.80} & \textbf{89.90} & \textbf{0.90} & \textbf{87.14} & \textbf{83.64} & \textbf{0.89} & \textbf{86.90} & \textbf{83.00} & \textbf{0.87} & \textbf{86.10} & \textbf{81.20} \\
\bottomrule
\end{tabular}
\vspace{-0.5ex}
\end{table*}

The ablation in Table~\ref{tab:ablation} corroborates this progression. Adding asymmetric priors (+0.8--1.1 QWK) mitigates misclassification of rare severe grades. Wasserstein alignment and AG-Soft further improve class-level balance by pulling clusters closer to an ordinal axis. The ORM head boosts macro-F1 through structured ordinal regression, while MAOC provides the most distinct leap, translating to sharper separation and improved QWK on all datasets. Notably, the full CAP-WAE surpasses prior generative or ordinal baselines not by marginal tuning, but by systematically integrating imbalance-, ordinality-, and geometry-aware design choices. These results justify the rationale behind each component, introduced to address a concrete limitation observed in prior baselines (imbalance, lack of ordinality, poor latent geometry), and their cumulative effect produces state-of-the-art performance that is consistent across heterogeneous DR benchmarks.

\begin{table*}[!ht]
\small
\centering
\caption{\textbf{Ablation study} of CAP-WAE across four DR benchmarks. We progressively add key components: asymmetric prior (AS), Wasserstein/MMD alignment, AG-Soft supervision, ORM head, and latent geometry priors (MAOC).}
\label{tab:ablation}
\setlength{\tabcolsep}{3pt}
\begin{tabular}{lccc|ccc|ccc|ccc}
\toprule
& \multicolumn{3}{c}{\textbf{Zenodo-DR-7} (7)} & \multicolumn{3}{c}{\textbf{APTOS-2019} (5)} & \multicolumn{3}{c}{\textbf{Messidor-2} (5)} & \multicolumn{3}{c}{\textbf{IDRiD} (5)} \\
\cmidrule(lr){2-4}\cmidrule(lr){5-7}\cmidrule(lr){8-10}\cmidrule(lr){11-13}
Variant & QWK & Acc & F1 & QWK & Acc & F1 & QWK & Acc & F1 & QWK & Acc & F1 \\
\midrule
VAE–KL    & 0.84 & 86.70 & 83.50 & 0.83 & 83.40 & 77.90 & 0.82 & 82.20 & 77.40 & 0.79 & 80.20 & 75.40 \\
\;\; + AS.\ (KL)      & 0.85 & 87.40 & 84.30 & 0.84 & 84.20 & 78.50 & 0.83 & 83.10 & 78.20 & 0.80 & 81.10 & 76.00 \\
WAE–MMD            & 0.86 & 88.00 & 84.90 & 0.85 & 85.00 & 79.50 & 0.84 & 83.60 & 79.00 & 0.81 & 81.60 & 76.50 \\
\;\; + AS.\ (MMD)     & 0.88 & 89.00 & 86.00 & 0.86 & 86.00 & 80.60 & 0.85 & 84.60 & 80.20 & 0.82 & 82.50 & 77.40 \\
\;\; + AG-Soft    & 0.89 & 89.50 & 86.60 & 0.87 & 86.50 & 81.10 & 0.86 & 85.30 & 81.00 & 0.83 & 83.20 & 78.20 \\
\;\; + ORM               & 0.90 & 90.00 & 87.20 & 0.88 & 86.90 & 81.70 & 0.87 & 85.80 & 81.50 & 0.84 & 83.70 & 78.80 \\
\;\; + MAOC        & 0.91 & 91.00 & 88.50 & 0.89 & 87.00 & 82.20 & 0.88 & 86.20 & 82.40 & 0.85 & 84.40 & 79.50 \\
\textbf{CAP-WAE}  & \textbf{0.94} & \textbf{91.80} & \textbf{89.90} & \textbf{0.90} & \textbf{87.14} & \textbf{83.64} & \textbf{0.89} & \textbf{86.90} & \textbf{83.00} & \textbf{0.87} & \textbf{86.10} & \textbf{81.20} \\
\bottomrule
\end{tabular}
\vspace{-0.5ex}
\end{table*}

\subsection{Qualitative Evaluation}

\begin{figure*}[t]
  \centering
  \includegraphics[width=\textwidth]{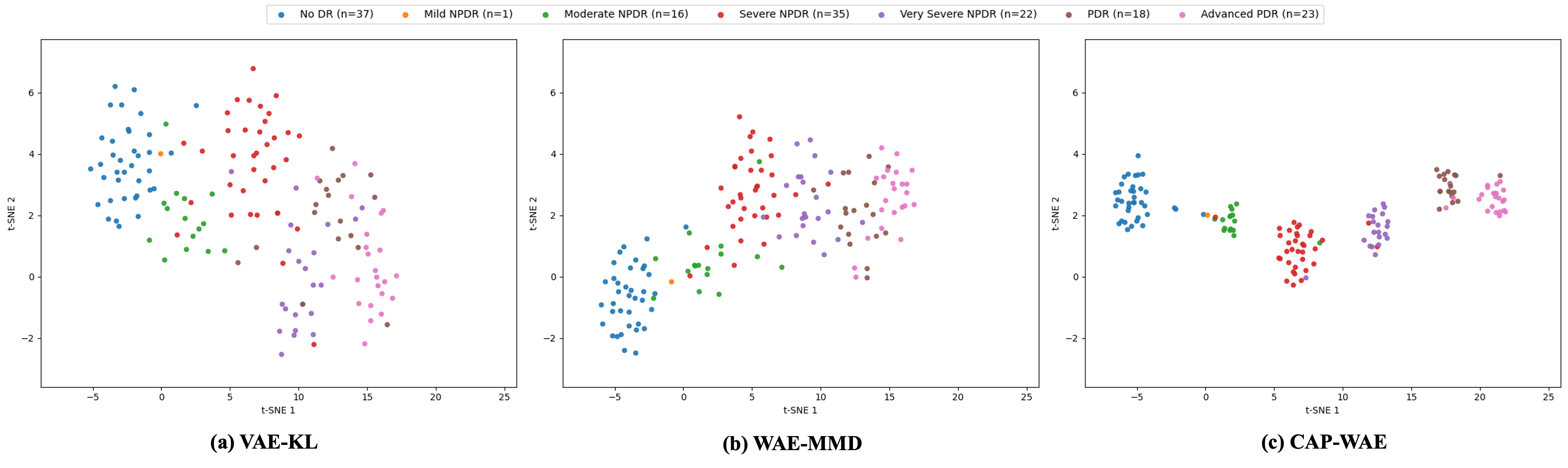}
  \caption{t-SNE latent-space evolution across ablation.
  Left: \emph{VAE–KL} baseline shows overlapping clusters and weak ordinal structure.
  Middle: \emph{WAE–MMD} improves separation but retains cross-class mixing.
  Right: \emph{Full CAP–WAE (ours)} exhibits compact, well-ordered bands along disease severity with clearer orthogonal separation.}
  \label{fig:tsne-ablation}
\end{figure*}

The t-SNE visualizations in Fig.~\ref{fig:tsne-ablation} provide qualitative evidence of how the latent space evolves across model variants. The baseline VAE–KL produces overlapping clusters with no clear ordering, reflecting its difficulty in modeling ordinal transitions. Introducing Wasserstein alignment (WAE–MMD) tightens clusters and yields partial separation, but cross-class mixing remains prevalent, consistent with the moderate gains in QWK and macro-F1. In contrast, the full CAP–WAE arranges clusters into compact, nearly disjoint bands that follow the severity progression from No DR through Advanced PDR. This ordered, orthogonally separated structure directly corresponds to the effect of the MAOC prior and ORM head, which explicitly encourage severity-aware alignment in the latent space. The qualitative alignment of visual cluster geometry with ordinal disease progression validates the quantitative improvements, confirming that CAP–WAE does not merely boost classification metrics but reshapes the latent representation into a semantically meaningful space.

\section{Conclusion}
\label{sec:conclusion}

We introduced CAP–WAE, a generative–discriminative framework that unifies constrained asymmetric priors, adaptive ordinal supervision, and manifold-aware latent regularization to jointly address class imbalance, disease severity ordering, and clinical asymmetry in diabetic retinopathy grading. Across four benchmarks, CAP–WAE consistently outperforms prior baselines in QWK, accuracy, and macro-F1, while t-SNE analyses show a progression from entangled to well-ordered, semantically meaningful latent manifolds. Despite these advances, the approach assumes stationarity of training distributions when fixing asymmetric priors, and the manifold orthogonality constraint (MAOC) enforces local alignment without theoretical guarantees of global separability. Moreover, performance depends on reliable ordinal annotations, which in clinical practice can be noisy or inconsistent. Future work should investigate adaptive prior learning under domain shift, integrate self-supervised or multi-modal representations to reduce annotation dependence, and extend to longitudinal modeling of progression. Beyond DR, the proposed framework offers a general template for ordinal, imbalanced medical grading tasks (e.g., cancer staging, fibrosis scoring), with the potential to mitigate under-grading errors and improve triage decisions in clinical workflows.

\bibliographystyle{abbrvnat}   
\bibliography{references}

\begin{thebibliography}{30}
\providecommand{\natexlab}[1]{#1}
\providecommand{\url}[1]{\texttt{#1}}
\expandafter\ifx\csname urlstyle\endcsname\relax
  \providecommand{\doi}[1]{doi: #1}\else
  \providecommand{\doi}{doi: \begingroup \urlstyle{rm}\Url}\fi

\bibitem[Abr{\`a}moff et~al.(2013)Abr{\`a}moff, Folk, Han, Walker, Williams, Russell, Massin, Cochener, Gain, Tang, et~al.]{abramoff2013automated}
M.~D. Abr{\`a}moff, J.~C. Folk, D.~P. Han, J.~D. Walker, D.~F. Williams, S.~R. Russell, P.~Massin, B.~Cochener, P.~Gain, L.~Tang, et~al.
\newblock Automated analysis of retinal images for detection of referable diabetic retinopathy.
\newblock \emph{JAMA ophthalmology}, 131\penalty0 (3):\penalty0 351--357, 2013.

\bibitem[{Asia Pacific Tele-Ophthalmology Society (APTOS)}(2019)]{aptos2019}
{Asia Pacific Tele-Ophthalmology Society (APTOS)}.
\newblock Aptos 2019 blindness detection.
\newblock \url{https://www.kaggle.com/competitions/aptos2019-blindness-detection}, 2019.
\newblock Kaggle competition dataset, accessed 2025-08-17.

\bibitem[Ben{\'\i}tez et~al.(2021)Ben{\'\i}tez, Matto, Rom{\'a}n, Noguera, Garc{\'\i}a-Torres, Ayala, Pinto-Roa, Gardel-Sotomayor, Facon, and Grillo]{benitez2021dataset}
V.~E.~C. Ben{\'\i}tez, I.~C. Matto, J.~C.~M. Rom{\'a}n, J.~L.~V. Noguera, M.~Garc{\'\i}a-Torres, J.~Ayala, D.~P. Pinto-Roa, P.~E. Gardel-Sotomayor, J.~Facon, and S.~A. Grillo.
\newblock Dataset from fundus images for the study of diabetic retinopathy.
\newblock \emph{Data in brief}, 36:\penalty0 107068, 2021.

\bibitem[Bodapati et~al.(2021{\natexlab{a}})Bodapati, Shaik, and Naralasetti]{bodapati2021composite}
J.~D. Bodapati, N.~S. Shaik, and V.~Naralasetti.
\newblock Composite deep neural network with gated-attention mechanism for diabetic retinopathy severity classification.
\newblock \emph{Journal of Ambient Intelligence and Humanized Computing}, 12\penalty0 (10):\penalty0 9825--9839, 2021{\natexlab{a}}.

\bibitem[Bodapati et~al.(2021{\natexlab{b}})Bodapati, Shaik, and Naralasetti]{bodapati2021deep}
J.~D. Bodapati, N.~S. Shaik, and V.~Naralasetti.
\newblock Deep convolution feature aggregation: an application to diabetic retinopathy severity level prediction.
\newblock \emph{Signal, Image and Video Processing}, 15\penalty0 (5):\penalty0 923--930, 2021{\natexlab{b}}.

\bibitem[Cao et~al.(2019)Cao, Wei, Gaidon, Arechiga, and Ma]{cao2019ldam}
K.~Cao, C.~Wei, A.~Gaidon, N.~Arechiga, and T.~Ma.
\newblock Learning imbalanced datasets with label-distribution-aware margin loss.
\newblock In \emph{Advances in Neural Information Processing Systems (NeurIPS)}, volume~32, 2019.
\newblock URL \url{https://proceedings.neurips.cc/paper/2019/hash/621461af90cadfdaf0e8d4cc25129f91-Abstract.html}.

\bibitem[Cao et~al.(2021)Cao, Mirjalili, and Raschka]{cao2021corn}
W.~Cao, V.~Mirjalili, and S.~Raschka.
\newblock Rank consistent ordinal regression for neural networks with application to age estimation.
\newblock In \emph{Proceedings of the AAAI Conference on Artificial Intelligence}, volume 35-2, pages 1062--1070, 2021.
\newblock URL \url{https://ojs.aaai.org/index.php/AAAI/article/view/16177}.

\bibitem[Castiglioni et~al.(2021)Castiglioni, Rundo, Codari, Di~Leo, Salvatore, Interlenghi, Gallivanone, Cozzi, D'Amico, and Sardanelli]{castiglioni2021ai}
I.~Castiglioni, L.~Rundo, M.~Codari, G.~Di~Leo, C.~Salvatore, M.~Interlenghi, F.~Gallivanone, A.~Cozzi, N.~C. D'Amico, and F.~Sardanelli.
\newblock Ai applications to medical images: From machine learning to deep learning.
\newblock \emph{Physica medica}, 83:\penalty0 9--24, 2021.

\bibitem[Chen et~al.(2022)Chen, Wang, Zhang, Fung, Thai, Moore, Mannel, Liu, Zheng, and Qiu]{chen2022recent}
X.~Chen, X.~Wang, K.~Zhang, K.-M. Fung, T.~C. Thai, K.~Moore, R.~S. Mannel, H.~Liu, B.~Zheng, and Y.~Qiu.
\newblock Recent advances and clinical applications of deep learning in medical image analysis.
\newblock \emph{Medical image analysis}, 79:\penalty0 102444, 2022.

\bibitem[Cherukuri et~al.(2024)Cherukuri, Shaik, and Ye]{cherukuri2024guided}
T.~K. Cherukuri, N.~S. Shaik, and D.~H. Ye.
\newblock Guided context gating: Learning to leverage salient lesions in retinal fundus images.
\newblock In \emph{2024 IEEE International Conference on Image Processing (ICIP)}, pages 3098--3104. IEEE, 2024.

\bibitem[Dosovitskiy et~al.(2021)Dosovitskiy, Beyer, Kolesnikov, Weissenborn, Zhai, Unterthiner, Dehghani, Minderer, Heigold, Gelly, Uszkoreit, and Houlsby]{dosovitskiy2021vit}
A.~Dosovitskiy, L.~Beyer, A.~Kolesnikov, D.~Weissenborn, X.~Zhai, T.~Unterthiner, M.~Dehghani, M.~Minderer, G.~Heigold, S.~Gelly, J.~Uszkoreit, and N.~Houlsby.
\newblock An image is worth 16x16 words: Transformers for image recognition at scale.
\newblock In \emph{International Conference on Learning Representations (ICLR)}, 2021.
\newblock URL \url{https://openreview.net/forum?id=YicbFdNTTy}.

\bibitem[Elguebaly and Bouguila(2013)]{elguebaly2013finite}
T.~Elguebaly and N.~Bouguila.
\newblock Finite asymmetric generalized gaussian mixture models learning for infrared object detection.
\newblock \emph{Computer Vision and Image Understanding}, 117\penalty0 (12):\penalty0 1659--1671, 2013.

\bibitem[Gretton et~al.(2012)Gretton, Borgwardt, Rasch, Sch{\"o}lkopf, and Smola]{Gretton2012MMD}
A.~Gretton, K.~M. Borgwardt, M.~J. Rasch, B.~Sch{\"o}lkopf, and A.~J. Smola.
\newblock A kernel two-sample test.
\newblock \emph{Journal of Machine Learning Research}, 13:\penalty0 723--773, 2012.

\bibitem[He et~al.(2020)He, Li, Li, Wang, and Fu]{he2020cabnet}
A.~He, T.~Li, N.~Li, K.~Wang, and H.~Fu.
\newblock Cabnet: Category attention block for imbalanced diabetic retinopathy grading.
\newblock \emph{IEEE Transactions on Medical Imaging}, 40\penalty0 (1):\penalty0 143--153, 2020.

\bibitem[Kingma and Welling(2014)]{Kingma2014VAE}
D.~P. Kingma and M.~Welling.
\newblock Auto-encoding variational bayes.
\newblock In \emph{International Conference on Learning Representations (ICLR)}, 2014.

\bibitem[Li et~al.(2021)Li, Bo, Hu, Kang, Liu, Wang, and Fu]{li2021applications}
T.~Li, W.~Bo, C.~Hu, H.~Kang, H.~Liu, K.~Wang, and H.~Fu.
\newblock Applications of deep learning in fundus images: A review.
\newblock \emph{Medical Image Analysis}, 69:\penalty0 101971, 2021.

\bibitem[Lin et~al.(2025)Lin, Nie, Wu, Shen, and Yang]{lin2025g2c}
B.~Lin, D.~Nie, X.~Wu, X.~Shen, and C.~Yang.
\newblock G2c-net: Grade-skewed domain adaptation network with coordinate and category attention for diabetic retinopathy grading.
\newblock \emph{Biomedical Signal Processing and Control}, 110:\penalty0 108203, 2025.

\bibitem[Ma et~al.(2025)Ma, Gu, Guo, Qin, Wen, Shi, Dai, and Chen]{Ma2025ABiD}
Y.~Ma, Y.~Gu, S.~Guo, X.~Qin, S.~Wen, N.~Shi, W.~Dai, and Y.~Chen.
\newblock Grade-skewed domain adaptation via asymmetric bi-classifier discrepancy minimization for diabetic retinopathy grading.
\newblock \emph{IEEE Transactions on Medical Imaging}, 44\penalty0 (3):\penalty0 1115--1129, Mar. 2025.
\newblock \doi{10.1109/TMI.2024.3485064}.

\bibitem[Porwal et~al.(2018)Porwal, Pachade, Kamble, Kokare, Deshmukh, Sahasrabuddhe, and Meriaudeau]{porwal2018indian}
P.~Porwal, S.~Pachade, R.~Kamble, M.~Kokare, G.~Deshmukh, V.~Sahasrabuddhe, and F.~Meriaudeau.
\newblock Indian diabetic retinopathy image dataset (idrid): a database for diabetic retinopathy screening research.
\newblock \emph{Data}, 3\penalty0 (3):\penalty0 25, 2018.

\bibitem[Rezende et~al.(2014)Rezende, Mohamed, and Wierstra]{Rezende2014Stochastic}
D.~J. Rezende, S.~Mohamed, and D.~Wierstra.
\newblock Stochastic backpropagation and approximate inference in deep generative models.
\newblock In \emph{International Conference on Machine Learning (ICML)}, pages 1278--1286, 2014.

\bibitem[Sakai(2022)]{sakai2022variants}
T.~Sakai.
\newblock On variants of root normalised order-aware divergence and a divergence based on kendall's tau.
\newblock \emph{arXiv preprint arXiv:2204.07304}, 2022.

\bibitem[Shaik and Cherukuri(2021)]{shaik2021lesion}
N.~S. Shaik and T.~K. Cherukuri.
\newblock Lesion-aware attention with neural support vector machine for retinopathy diagnosis.
\newblock \emph{Machine Vision and Applications}, 32\penalty0 (6):\penalty0 126, 2021.

\bibitem[Shaik and Cherukuri(2022)]{shaik2022hinge}
N.~S. Shaik and T.~K. Cherukuri.
\newblock Hinge attention network: A joint model for diabetic retinopathy severity grading.
\newblock \emph{Applied Intelligence}, 52\penalty0 (13):\penalty0 15105--15121, 2022.

\bibitem[Simonyan and Zisserman(2014)]{simonyan2014very}
K.~Simonyan and A.~Zisserman.
\newblock Very deep convolutional networks for large-scale image recognition.
\newblock \emph{arXiv preprint arXiv:1409.1556}, 2014.

\bibitem[Tang et~al.(2023)Tang, Yang, and Song]{tang2023disease}
W.~Tang, Z.~Yang, and Y.~Song.
\newblock Disease-grading networks with ordinal regularization for medical imaging.
\newblock \emph{Neurocomputing}, 545:\penalty0 126245, 2023.

\bibitem[Tolstikhin et~al.(2018)Tolstikhin, Bousquet, Gelly, and Sch{\"o}lkopf]{Tolstikhin2018WAE}
I.~O. Tolstikhin, O.~Bousquet, S.~Gelly, and B.~Sch{\"o}lkopf.
\newblock Wasserstein auto-encoders.
\newblock In \emph{International Conference on Learning Representations (ICLR)}, 2018.

\bibitem[Vargas et~al.(2025)Vargas, Guti{\'e}rrez, Rosati, Romeo, Frontoni, and Herv{\'a}s-Mart{\'\i}nez]{Vargas2025AGAsymGrading}
V.~M. Vargas, P.~A. Guti{\'e}rrez, R.~Rosati, L.~Romeo, E.~Frontoni, and C.~Herv{\'a}s-Mart{\'\i}nez.
\newblock Disease-grading networks with asymmetric gaussian distribution for medical imaging.
\newblock \emph{IEEE Transactions on Medical Imaging}, 2025.
\newblock Accepted; early access DOI: 10.1109/TMI.2025.3575402.

\bibitem[Wang et~al.(2023)Wang, Li, Huang, Cao, He, and He]{Wang2023L2RCLIP}
R.~Wang, P.~Li, H.~Huang, C.~Cao, R.~He, and Z.~He.
\newblock Learning-to-rank meets language: Boosting language-driven ordering alignment for ordinal classification.
\newblock In \emph{Advances in Neural Information Processing Systems (NeurIPS)}, 2023.
\newblock L2RCLIP preprint/code: https://github.com/raywang335/L2RCLIP.

\bibitem[Wen et~al.(2023)Wen, Zhang, Yao, and Yang]{Wen2023OLDL}
C.~Wen, X.~Zhang, X.~Yao, and J.~Yang.
\newblock Ordinal label distribution learning.
\newblock In \emph{Proceedings of the IEEE/CVF International Conference on Computer Vision (ICCV)}, pages 23481--23491, 2023.

\bibitem[Wilkinson et~al.(2003)Wilkinson, Ferris~III, Klein, Lee, Agardh, Davis, Dills, Kampik, Pararajasegaram, Verdaguer, et~al.]{wilkinson2003proposed}
C.~P. Wilkinson, F.~L. Ferris~III, R.~E. Klein, P.~P. Lee, C.~D. Agardh, M.~Davis, D.~Dills, A.~Kampik, R.~Pararajasegaram, J.~T. Verdaguer, et~al.
\newblock Proposed international clinical diabetic retinopathy and diabetic macular edema disease severity scales.
\newblock \emph{Ophthalmology}, 110\penalty0 (9):\penalty0 1677--1682, 2003.

\end{thebibliography}

\end{document}